\documentclass[journal]{IEEEtran}
\usepackage[noadjust]{cite}
\usepackage{srcltx}
\usepackage{graphicx}
\usepackage{epstopdf}
\usepackage{psfrag}
\usepackage{epsfig}
\usepackage[tbtags,sumlimits,nointlimits,reqno]{amsmath}
\usepackage{amssymb}
\usepackage{xcolor}
\usepackage{psfrag}
\usepackage{epstopdf}
\usepackage{pstool}
\usepackage[position=t,singlelinecheck=off]{subfig}

\usepackage{soul}
\usepackage{footnote}
\usepackage{amsmath}
\usepackage{color}
\usepackage{tcolorbox}
\usepackage{titling}

\newcommand{\ve}[1]{\boldsymbol{#1}}
\newcommand{\te}[1]{\overline{\overline{#1}}}
\tcbset{colframe=blue!100!,colback=blue!10!,,colupper=blue!100!}

\begin{document}
\pretitle{\begin{flushleft}\Huge}
\title{\textsf{Nonreciprocal Nongyrotropic Magnetless Metasurface}}
\posttitle{\par\end{flushleft}}
\vspace{2 mm}
\preauthor{\begin{flushleft}\large \lineskip 0.5em}
\author{\textsf{\textbf{{\Large Sajjad Taravati$^1$, Bakhtiar A. Khan$^2$, Shulabh Gupta$^3$, Karim Achouri$^1$,\\ and Christophe Caloz$^1$}}\\
\vspace{2 mm}
{\normalsize $^1$Department
of Electrical Engineering and Poly-Grames research center, \'{E}cole Polytechnique de Montr\'{e}al, Montr\'{e}al, Qu\'{e}bec, Canada\\
$^2$Department of Electrical and Computer Engineering, Concordia University, Montr\'{e}al, Qu\'{e}bec, Canada\\
$^3$Department of Electronics, Carleton University, Ottawa, Ontario, Canada}}}
\postauthor{\par\end{flushleft}}


\maketitle

\begin{abstract}
\textsf{We introduce a nonreciprocal nongyrotropic magnetless metasurface. In contrast to previous nonreciprocal structures, this metasurface does not require a biasing magnet, and is therefore lightweight and amenable to integrated circuit fabrication. Moreover, it does not induce Faraday rotation, and hence does not alter the polarization of waves, which is a desirable feature in many nonreciprocal devices. The metasurface is designed according to a Surface-Circuit-Surface (SCS) architecture and leverages the inherent unidirectionality of transistors for breaking time reversal symmetry. Interesting features include transmission gain as well as broad operating bandwidth and angular sector operation. It is finally shown that the metasurface is bianisotropic in nature, with nonreciprocity due to the electric-magnetic coupling parameters, and structurally equivalent to a moving uniaxial metasurface.}
\end{abstract}

Over the past decade, metasurfaces have spurred huge interest in the scientific community due to their unique optical properties~\cite{holloway2012overview,Capasso_NM_2014,AchouriEPJAM_2015}. Metasurfaces may be seen as the two-dimensional counterparts of volume metamaterials~\cite{Pendry_Sc_2004,caloz2005electromagnetic,Chen_Na_2006,engheta2006metamaterials,Shalaev_NP_2007,capolino2009theory,Zayats_Wiley_2013}, where Snell's law is generalized by the introduction of an abrupt phase shift along the optical path, leading to effects such as anomalous reflection and refraction of light~\cite{Capasso_Sc_2011} and a diversity of unprecedented wave transformation functionalities~\cite{Cappaso_JSQ_2013}.

The vast majority of the metasurfaces reported to date are restricted to reciprocal responses. Introducing nonreciprocity requires breaking time reversal symmetry. This can be accomplished via the magneto-optical effect~\cite{Gurevich_1996,Fan_OL_2005,Fan_PRL_2008,Ross_NP_2011,Parsa_TAP_03_2011,Steinberg_PRL_2014}, nonlinearity~\cite{Engheta_NC_2014,Valev_AM_2014,Gu_SR_2016}, space-time modulation~\cite{Fan_NPH_2009,Sounas_NC_2013,Alu_PRB_2015,Shalaev_OME_2015,Fan_APL_2016,Alu_PNAS_2016,Taravati_LWA_2016_arXiv} or metamaterial transistor loading~\cite{kodera2011artificial,Joannopoulos_PNAS_2012,Kodera_TMTT_03_2013,sounas2013electromagnetic}. However, all these approaches suffer from a number of drawbacks. The magneto-optical approach requires bulky, heavy and costly magnets~\cite{Fan_PRL_2008}. The nonlinear approach involves dependence to signal intensity and severe nonreciprocity-loss trade-off~\cite{Fan_NP_2015}. The space-time modulation approach implies high design complexity, especially for a spatial device such as a metasurface. Finally, the transistor-based nonreciprocal metasurfaces reported in~\cite{Joannopoulos_PNAS_2012,sounas2013electromagnetic} are intended to operate as Faraday rotators, whereas gyrotropy is undesired in applications requiring nonreciprocity without alteration of the wave polarization, such as for instance one-way screens, isolating radomes, radar absorbers or illusion cloaks.

We introduce here the concept of a nonreciprocal nongyrotropic magnetless metasurface and demonstrate a simple three-layer Surface-Circuit-Surface (SCS) implementation of it. In the proposed metasurface, time reversal symmetry is broken by the presence of unilateral transistors in the circuit part of the SCS structure, which is appropriate in the microwave and millimeter-wave regime. A space-time modulated version of the structure, although nontrivial, may also be envisioned for the optical regime. The metasurface is shown to work for all incidence angles and to provide gain. It is finally shown to be structurally equivalent to a moving uniaxial metasurface.

\vspace{2 mm}
\begin{tcolorbox}
\textsf{\textbf{Significance Statement}\\[2mm]
 While most materials and metamaterials are reciprocal, i.e. characterized by symmetric scattering parameters, nonreciprocal devices (e.g. isolators, circulators, nonreciprocal phase shifter and polarizers) play a fundamental role in a great diversity of microwave and optical applications. Achieving nonreciprocity requires breaking the time reversal symmetry of a system using an external ``force’’, such as an externally applied field with a specific (bias) direction. Recently, magnet-less nonreciprocal metamaterials have been reported as advantageous alternatives to ferromagnetic materials. We introduce here a magnet-less nonreciprocal metasurface that is immune of Faraday rotation and addresses therefore a novel range of potential applications, that may include for instance novel one-way screens, isolating radomes, radar absorbers and illusion cloaks.}
\end{tcolorbox}

\section*{\textbf{\textsf{Operation Principle}}}
\begin{figure}
\begin{center}
\includegraphics[width=1\columnwidth]{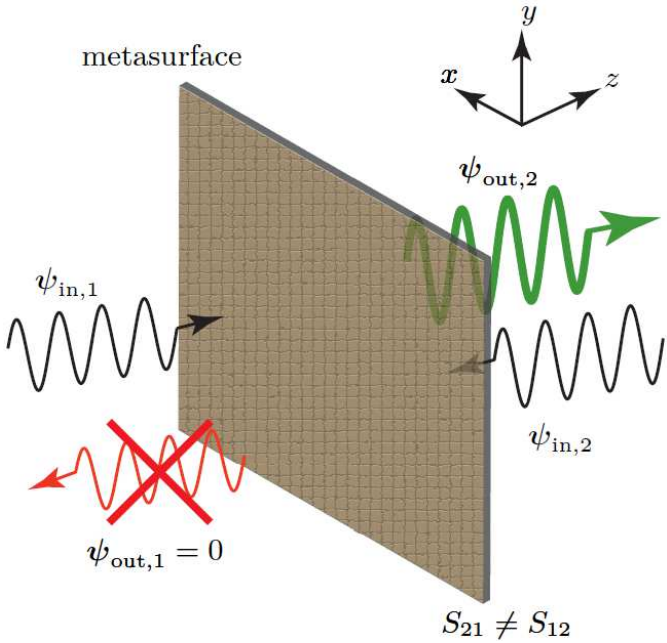}
\caption{Nonreciprocal nongyrotropic metasurface functionality.}\label{Fig:NRNG_gen}
\end{center}
\end{figure}

Figure~\ref{Fig:NRNG_gen} depicts the functionality of the nonreciprocal nongyrotropic metasurface. A wave traveling along the $+z$ direction, $\ve{\psi} _\text{in,1}$, passes through the metasurface, possibly with gain, without polarization alteration, from side~1 to side~2. In contrast, a wave traveling along the opposite direction from side~2, $\ve{\psi} _\text{in,2}$, is being absorbed and reflected (still without polarization alteration) by the metasurface and can not pass through the metasurface from side~2 to side~1. Such a metasurface is nonreciprocal, and may hence be characterized by asymmetric scattering parameters, $S_{21} \ne S_{12}$, where $S_{21}=\ve{\psi} _\text{out,2}/\ve{\psi} _\text{in,1} >1$ and $S_{12}=\ve{\psi} _\text{out,1}/\ve{\psi} _\text{in,2} <1$. Moreover, the metasurface is nongyrotropic since it does not induce any rotation of the incident electromagnetic field.

\begin{figure}[!ht]
\begin{center}
{\psfrag{a}[c][c][0.9][0]{\shortstack{metasurface}}
\psfrag{d}[c][c][0.9][17]{\shortstack{circuit}}
\psfrag{c}[l][c][1]{\shortstack{$\delta << \lambda$}}
\psfrag{e}[c][c][1]{$\ve{\psi} _\text{in}$}
\psfrag{f}[c][c][1]{$\ve{\psi} _\text{out}$}
\psfrag{g}[c][c][0.9][-28]{surface}
\psfrag{h}[c][c][0.9][-28]{surface}
\psfrag{x}[c][c][1]{$x$}
\psfrag{y}[c][c][1]{$y$}
\psfrag{z}[c][c][1]{$z$}
\includegraphics[width=0.8\columnwidth]{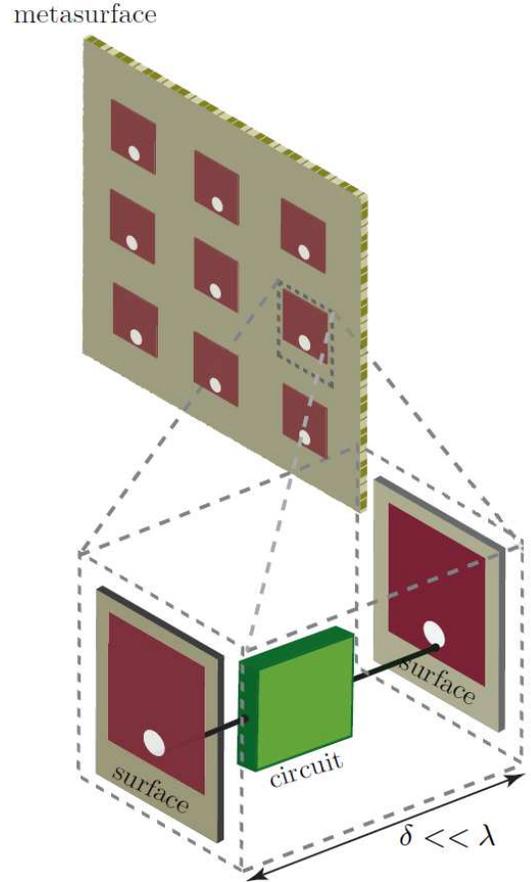}}
\caption{Surface-Circuit-Surface (SCS) metasurface architecture for the magnetless implementation of the nonreciprocal nongyrotropic metasurface in Fig.~\ref{Fig:NRNG_gen}.}\label{Fig:NRNG_imp}
\end{center}
\end{figure}

 To realize such a nonreciprocal and nongyrotropic metasurface, we employ the three-layer Surface-Circuit-Surface (SCS) architecture represented in Fig.~\ref{Fig:NRNG_imp}. The first surface receives the incoming wave from one side of the metasurface and feeds it into the circuit while the second surface collects the wave exiting the circuit and radiates it to the other side of the metasurface. The metasurface is constituted of an array of unit cells, themselves composed of two subwavelengthly spaced microstrip patch antennas interconnected by the circuit that will introduce transmission gain in one direction and transmission loss in the other direction.

\begin{figure*}[!ht]
\begin{center}
\includegraphics[width=2\columnwidth]{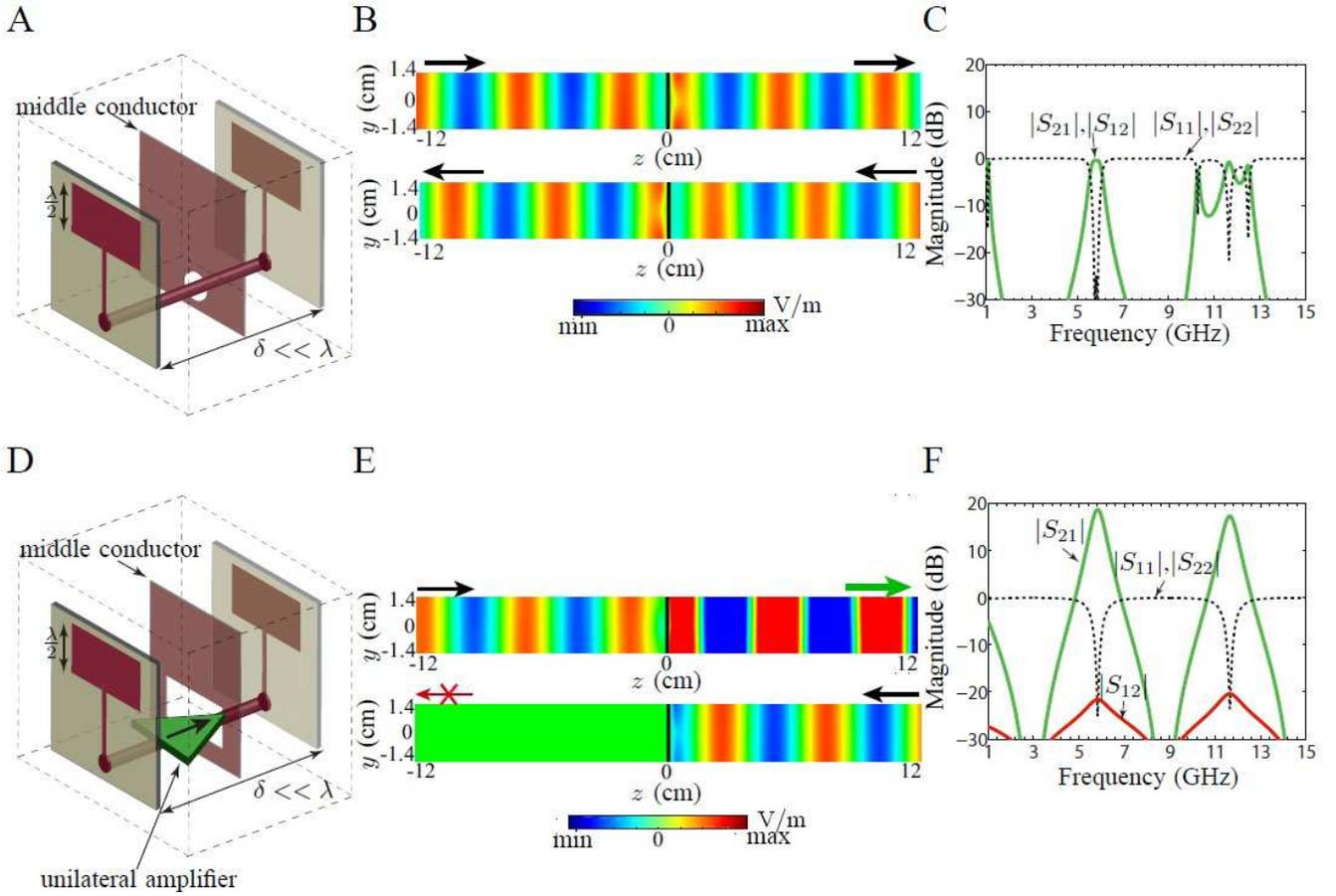}
\caption{Unit cell of the metasurface in Fig.~\ref{Fig:NRNG_imp} with (A,B,C) a direct connection for the circuit, corresponding to a reciprocal metasurface, and (D,E,F) a unilateral device (typically a transistor) for the circuit, corresponding to a nonreciprocal metasurface. (A,D) Structure. (B,E) Full-wave (FDTD) electric field distribution for excitations from the left and right (bottom). (C,F) Full-wave (FDTD) scattering parameter magnitudes.}
\label{Fig:unit_cell_T}
\end{center}
\end{figure*}

To best understand the impact of the circuit on the metasurface functionality, first consider the reciprocal unit cell of Fig.~\ref{Fig:unit_cell_T}A, where the interconnecting circuit is a direct connection (simple conducting wire). A conducting sheet is placed between the two patches to prevent any interaction between them. Figures~\ref{Fig:unit_cell_T}B and~\ref{Fig:unit_cell_T}C show the Finite Difference Time Domain (FDTD) response of structure in Fig.~\ref{Fig:unit_cell_T}A. Figure~\ref{Fig:unit_cell_T}B plots the electric field distribution for wave incidence from the left and right, with the metasurface being placed at $z=0$. The response is obviously reciprocal. The corresponding pass-bands are apparent in the scattering parameter magnitudes plotted in Fig.~\ref{Fig:unit_cell_T}C.

Consider now the nonreciprocal unit cell of Fig.~\ref{Fig:unit_cell_T}D, where the interconnecting circuit is a unilateral device, typically a transistor-based amplifier. Figures~\ref{Fig:unit_cell_T}E and~\ref{Fig:unit_cell_T}F show the corresponding FDTD response. Figure~\ref{Fig:unit_cell_T}E plots the electric field distribution for wave incidence from the left and right. When the excitation is from the left, the incoming wave passes through the structure, where it also gets amplified, and radiates to the right of the metasurface. When the excitation is from the right, the incoming wave is blocked, namely absorbed and reflected, by the metasurface. The pass-band ($S_{11}\simeq 0$) and stop-band ($S_{11}\simeq 1$) are shown in in Fig.~\ref{Fig:unit_cell_T}F. The explanation of the multiple pass-bands suppression is provided in the supporting information section.
\section*{\textbf{\textsf{Experimental Demonstration}}}
\begin{figure*}
\begin{center}
\includegraphics[width=2\columnwidth]{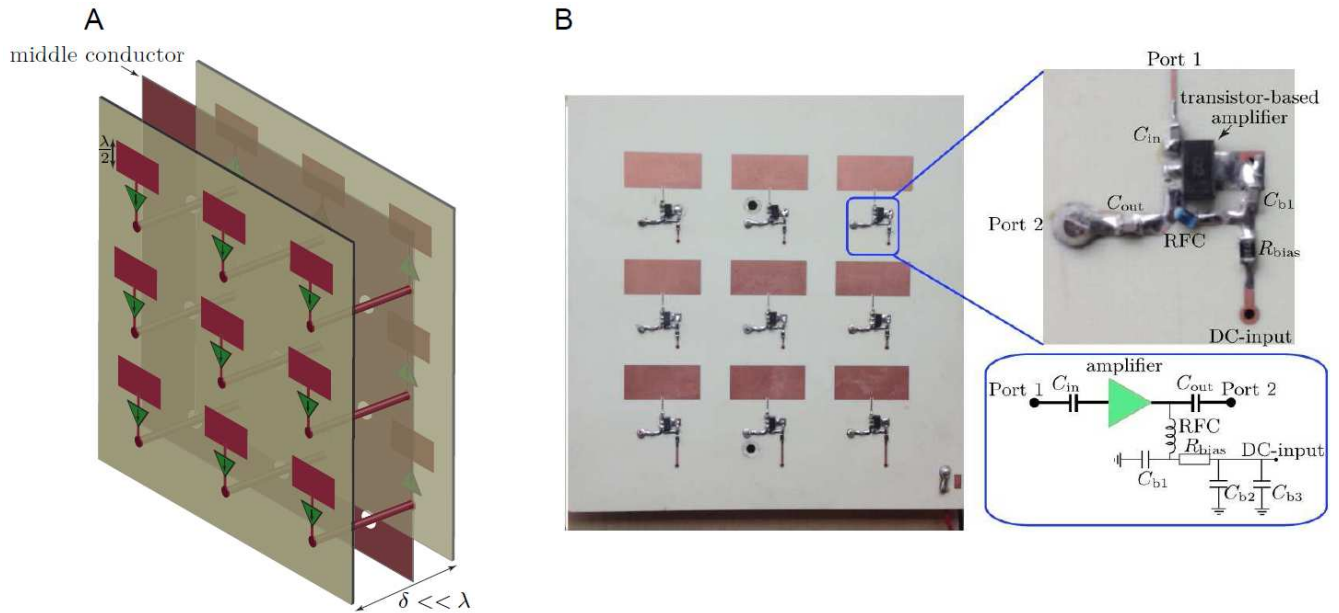}
\caption{Realized $3\times 3$-cell implementation of the metasurface, where, compared to Fig.~\ref{Fig:unit_cell_T}D, the transistors have been shifted to the surfaces for fabrication convenience. (A)~Exploded perspective view. (B)~Photograph with zoom on transistor part and corresponding biasing network.}
\label{Fig:fabric_struc}
\end{center}
\end{figure*}

Figure~\ref{Fig:fabric_struc} shows the realized $3\times 3$ metasurface, based on the SCS architecture of Fig.~\ref{Fig:unit_cell_T}D. The metasurface is designed to operate in the frequency range from 5.8 to 6~GHz. Its thickness is deeply subwavelength, specifically $\delta\approx\lambda_0/30$, where $\lambda_0$ is the wavelength at the center frequency, $5.9$~GHz, of the operating frequency range. More details about the fabricated metasurface are provided in the Materials and Methods section.

\begin{figure}
\begin{center}
{
\psfrag{A}[c][c][0.9]{Frequency (GHz)}
\psfrag{B}[c][c][0.9]{Transmission Magnitude (dB)}
\psfrag{C}[c][c][0.8]{$S_{21}$}
\psfrag{D}[c][c][0.8]{$S_{12}$}
\psfrag{E}[l][c][0.8]{isolation}
\psfrag{a}[l][c][0.8]{metasurface}
\psfrag{b}[c][c][0.75]{antenna 1}
\psfrag{c}[c][c][0.75]{antenna 2}
\includegraphics[width=0.8\columnwidth]{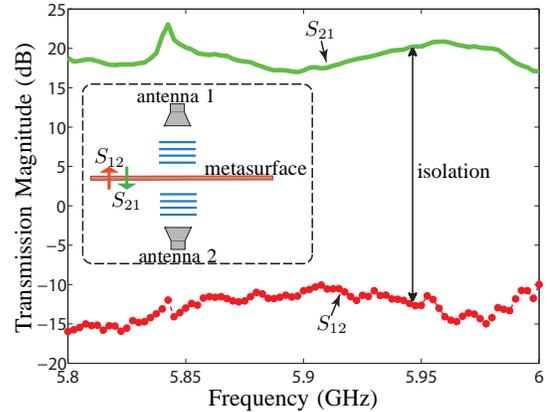}}
\caption{Experimental scattering parameters versus frequency for normal incidence and transmission.}
\label{Fig:result_norm}
\end{center}
\end{figure}

Figure~\ref{Fig:result_norm} shows the measured transmission scattering parameters versus frequency for normally aligned transmit and receive antennas. In the $1\rightarrow 2$ direction, more than $17$~dB transmission gain is achieved in the frequency range of interest, while in the $2\rightarrow 1$ direction, more than $10$~dB transmission loss is ensured across the same range, corresponding to an isolation of more than $27$~dB.

\begin{figure*}
\begin{center}
 \includegraphics[width=0.9\textwidth]{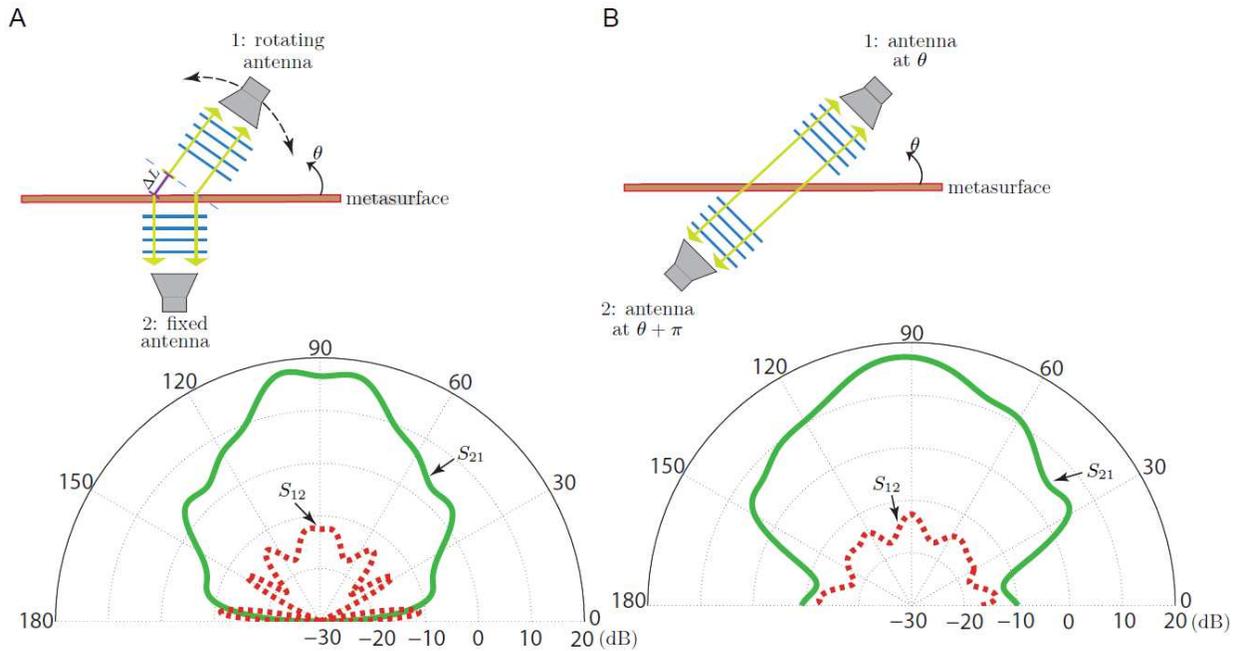}
  \caption{Experimental scattering parameters versus angle at $f=5.9$~GHz for transmission (A) under normal (one side) and oblique (other side) angles, (B) in a straight line under an oblique angle.}
  \label{Fig:meas_scan_resu}
  \end{center}
\end{figure*}

Two other experiments are next carried out to investigate the angular dependence of the metasurface response. In the first experiment, we fix the position of one antenna normal to the metasurface and rotate the other antenna from $0$ to $180^\circ$ with respect to the metasurface, as illustrated at the top of Fig.~\ref{Fig:meas_scan_resu}A. The bottom of Fig.~\ref{Fig:meas_scan_resu}A shows the measured transmission levels for both directions, $|S_{21}|$ and $|S_{12}|$. We observe that the metasurface passes the wave with gain over a beamwidth of about $110^\circ$ from $\theta=35^\circ$ to $145^\circ$ in the $1\rightarrow 2$ direction and attenuates it by more than $12$~dB in the $2\rightarrow 1$ direction, which corresponds to a minimum isolation of about $15$~dB across the aforementioned beamwidth.

In the next angular dependence experiment, we rotate two antennas rigidly aligned from $0$ to $180^\circ$ with respect to the metasurface, as illustrated at the top of Fig.~\ref{Fig:meas_scan_resu}B. The bottom of Fig.~\ref{Fig:meas_scan_resu}B shows the measured transmission levels for both directions. We observe that the metasurface passes the wave with gain over a beamwidth of about $130^\circ$ from $\theta=25^\circ$ to $155^\circ$ in the $1\rightarrow 2$ direction and attenuates it by more than $12$~dB in the $2\rightarrow 1$ direction, which corresponds to a minimum isolation of about $21$~dB across the aforementioned beamwidth.

Comparing Figs.~\ref{Fig:meas_scan_resu}A and~\ref{Fig:meas_scan_resu}B reveals that the transmission gain is less sensitive to angle in the latter case. This is due to the fact that the optical path difference between any pair of rays across the SCS structure is null when the antennas are aligned, as in Fig.~\ref{Fig:meas_scan_resu}B, whereas it is angle-dependent otherwise, as in Fig.~\ref{Fig:meas_scan_resu}A (path difference $\Delta L$).

The experimental results presented above show that the proposed nonreciprocal nongyrotropic metasurface works as expected, with remarkable efficiency. Moreover, it exhibits the following additional favorable features. Firstly, it provides gain, which makes it particularly efficient as a repeater device. Secondly, in contrast to other nonreciprocal metasurfaces, the structure is not limited to the monochromatic regime since patch antennas are fairly broadband and their bandwidth can be enhanced by various standard techniques~\cite{Garg_2001}. Note that the structure presented here has not been optimized in this sense but already features a bandwidth of over $3\%$. Thirdly, the metasurface exhibits a very wide operating angular sector, due to both aforementioned small or null optical path difference between different rays, due to the SCS architecture, and the inherent low directivity of patch antenna elements. The reported operating sectors, of over $100^\circ$, are much larger than those of typical metasurfaces.

\subsection*{\textbf{\textsf{Characterization as a Bianisotropic Metasurface}}}
The nonreciprocal nongyrotropic metasurface has been conceived from an engineering perspective in the previous sections. We present here an alternative theoretical approach, that will reveal the equivalence to a moving medium to be covered in the next section. Not knowing a priori the electromagnetic nature of the metasurface, we start from the most general case of a bianisotropic medium, characterized by the following spectral relations
\begin{subequations}
\label{Eq:ConstRel}
\begin{equation}
\ve{D} = \te{\epsilon} \cdot \ve{E} + \te{\xi}\cdot\ve{H},
\end{equation}
\begin{equation}
\ve{B} =  \te{\zeta}\cdot \ve{E} + \te{\mu}\cdot\ve{H}.
\end{equation}
\end{subequations}

A metasurface can be generally characterized by the following continuity equations,
\begin{subequations}
\label{Eq:MSBC}
\begin{equation}
\hat{z}\times\Delta\ve{H}
=j\omega\epsilon_0\te{\chi}_\text{ee}\cdot\ve{E}_\text{av}+jk_0\te{\chi}_\text{em}\cdot\ve{H}_\text{av},
\end{equation}
\begin{equation}
\Delta\ve{E}\times\hat{z}
=j\omega\mu_0 \te{\chi}_\text{mm}\cdot\ve{H}_\text{av}+jk_0\te{\chi}_\text{me}\cdot\ve{E}_\text{av},
\end{equation}
\end{subequations}
which relate the electromagnetic fields on both sides of a metasurface to its susceptibilities and assume here no normal susceptibility components~\cite{achouri2014general}. In these relations, $\Delta$ and the subscript `av' denote the difference of the fields and the average of the fields between both sides of the metasurface.

The susceptibilities in~\eqref{Eq:MSBC} are related to the constitutive parameters in~\eqref{Eq:ConstRel} as
\begin{subequations}
\label{eq:const_para_xi}
\begin{equation}
\te{\epsilon}=\epsilon_0(\te{I}+\te{\chi}_\text{ee}),\quad\te{\mu}=\mu_0(\te{I}+\te{\chi}_\text{mm}),
\end{equation}
\begin{equation}
\te{\xi}=\te{\chi}_\text{em}/c_0,\quad\te{\zeta}=\te{\chi}_\text{me}/c_0.
\end{equation}
\end{subequations}
Let us now solve the synthesis problem, which consists in finding the susceptibilities providing the nonreciprocal nongyrotropic response for the metasurface. This consists in substituting the electromagnetic fields of the corresponding transformation into~\eqref{Eq:MSBC}, as described in~\cite{achouri2014general}. Specifically, the transformation consists in passing a normally incident and forward propagating ($+z$) plane wave through the metasurface with transmission coefficient $T=1$ and fully absorbing a normally incident wave in the opposite direction. The result reads~\cite{AchouriEPJAM_2015}

\begin{subequations}
\label{eq:ChiTrans}
\begin{equation}
\label{eq:ChiTrans1}
\te{\chi}_\text{ee}= -\frac{j}{k_0}\begin{pmatrix} 1 & 0 \\ 0 & 1 \end{pmatrix},
\quad
\te{\chi}_\text{mm}= -\frac{j}{k_0}\begin{pmatrix} 1 & 0 \\ 0 & 1 \end{pmatrix},
\end{equation}
\begin{equation}
\label{eq:ChiTrans2}
\te{\chi}_\text{em}= \frac{j}{k_0}\begin{pmatrix} 0 & 1 \\ -1 & 0 \end{pmatrix},
\quad
\te{\chi}_\text{me}= \frac{j}{k_0}\begin{pmatrix} 0 & -1 \\ 1 & 0 \end{pmatrix},
\end{equation}
\end{subequations}
and reveals that nonreciprocity in the metasurface is due to the electric-magnetic coupling contributions, $\te{\chi}_\text{em}$ and $\te{\chi}_\text{me}$.
\subsection*{\textbf{\textsf{Equivalence with a Moving Metasurface}}}

The expressions in~\eqref{eq:ChiTrans} suggest an alternative implementation of the nonreciprocal nongyrotropic metasurface. Indeed, the form of these susceptibility tensors is identical to that of a moving uniaxial medium~\cite{Kong_1986}. Such a medium, assuming motion in the $z$-direction, is characterized by the tensor set
\begin{equation}
\label{Eq:MovingMed}
\begin{pmatrix}
\te{\epsilon}'  &  \te{\xi}' \\
\te{\zeta}' &  \te{\mu}'
\end{pmatrix}=
\begin{pmatrix}
\epsilon' &  0& 0 & 0& 0 & 0 \\
0 &  \epsilon'  &0 &  0 & 0 & 0 \\
0 &  0  &\epsilon_z &  0 & 0 & 0 \\
0 &  0 & 0  & \mu' & 0 & 0\\
0 &  0 & 0  & 0& \mu' & 0 \\
0 &  0 & 0  & 0& 0 & \mu_z
\end{pmatrix},
\end{equation}
where the primes denote the moving frame of reference and where $\epsilon_z$ and $\mu_z$ can take arbitrary values. To an observer in the rest frame of reference, this tensor set transforms to the bianisotropic set
\begin{equation}
\label{Eq:CTens}
\begin{pmatrix}
\te{\epsilon}  &  \te{\xi} \\
\te{\zeta} &  \te{\mu}
\end{pmatrix}=
\begin{pmatrix}
\epsilon &  0& 0 & 0& \xi & 0 \\
0 &  \epsilon  &0 &  -\xi & 0 & 0 \\
0 &  0  &\epsilon_z &  0 & 0 & 0 \\
0 &  -\xi & 0  & \mu & 0 & 0\\
\xi &  0 & 0  & 0& \mu & 0 \\
0 &  0 & 0  & 0& 0 & \mu_z
\end{pmatrix},
\end{equation}
whose elements are found using the Lorentz transform operation~\cite{Kong_1986}
\begin{equation}
\label{Eq:C}
\te{C} = \te{L}_6^{-1} \cdot \te{C}' \cdot \te{L}_6,
\end{equation}
where the matrices $\te{C}$ and $\te{L}_6$ are respectively given by
\begin{equation}
\label{Eq:CMat}
\te{C}=
\begin{pmatrix}
c(\te{\epsilon} - \te{\xi}\cdot \te{\mu}^{-1} \cdot \te{\zeta}) &  \te{\xi}\cdot\te{\mu}^{-1} \\
-\te{\mu}^{-1}\cdot \te{\zeta} &  \te{\mu}^{-1}/c
\end{pmatrix}
\end{equation}
and
\begin{equation}
\label{Eq:L6}
\te{L}_6 =\gamma
\begin{pmatrix}
1 &  0 & 0 & 0 &-\beta & 0 \\
0 &  1 & 0 & \beta & 0 & 0 \\
0 &  0 & 1/\gamma & 0  & 0 & 0 \\
0 &  \beta & 0 & 1 & 0 & 0 \\
-\beta &  0 & 0 & 0 & 1 & 0 \\
0 &  0 & 0 & 0 & 0 & 1/\gamma
\end{pmatrix},
\end{equation}
where $\gamma= 1/\sqrt{1-\beta^2}$, $\beta = v/c_0$, with $v$ the velocity of the medium and $c_0$ the speed of light in vacuum. \eqref{Eq:CTens} is indeed identical to~\eqref{eq:ChiTrans}.

From this point, one can find the moving uniaxial metasurface that is equivalent to the nonreciprocal nongrytropic metasurface. This is accomplished by inserting the specified susceptibilities in~\eqref{eq:ChiTrans} into~\eqref{eq:const_para_xi}, which provides the values in~\eqref{Eq:CTens}, and then solve~\eqref{Eq:C} for $\te{C}'$ and $v$. It may be easily verified that one finds $\epsilon_r'=\mu_r'=1$ and $v=c_0$. This may be interpreted as follows: the forward propagating wave would simply fully transmit through ``moving vacuum'' while the backward propagating wave would never be able to catch-up with it and, thus, never pass through. For $T\neq 1$, one would find a complex velocity. The fact that this design approach is practically impossible indicates that the engineering approach is clearly preferable.

\section*{\textbf{\textsf{Materials and Methods}}}

The metasurface was realized using multilayer circuit technology, where two $6.2$~in $\times$~$6.2$~in RO4350 substrates with thickness $h = 30$~mil were assembled to realize a three metallization layer structure. The permittivity of the substrates are $\epsilon=\epsilon_\text{r}(1-j \tan \delta)$, with $\epsilon_\text{r} = 3.66$ and $\tan \delta = 0.0037$ at 10 GHz. The middle conductor of the structure (Fig.~\ref{Fig:fabric_struc}A) both supports the DC feeding network of the amplifiers and acts as the RF ground plane for the patch antennas. The dimensions of the 2$\times$9 microstrip patches are $1.08$~in $\times$~$0.49$~in.

The connections between the layers are provided by an array of circular metalized via holes, with 18 vias of 30~mil diameter connecting the DC bias network to the amplifiers, while the ground reference for the amplifiers is ensured by 18~sets of 6~vias of 20 mils diameter with 60~mils spacing. The connection between the two sides of the metasurface is provided by 9 via holes (Fig.~\ref{Fig:fabric_struc}A), with optimized dimensions of $60$~mils for the via diameters, $105$~mils for the pad diameters and $183$~mils for the hole diameter in the via middle conductor.

For the unilateral components, we used 18 Mini-Circuits Gali-2+ Darlington pair amplifiers. The amplifier circuit is shown in Fig.~\ref{Fig:fabric_struc}B, where $C_\text{in}=$1~pF and $C_\text{out}=$1~pF are DC-block capacitors, and $C_\text{b1}=4.7$~pF, $C_\text{b2}=1$~nF and $C_\text{b3}=1$ are a set of AC by-pass capacitors. A 4.5-V DC-supply provides the DC signal for the amplifiers through the DC network with a bias resistor of $R_\text{bias}=39$~Ohm corresponding to a DC current of $40$~mA for each amplifier.

The measurements were performed by a 37369D Anritsu network analyzer where two microstrip array antennas were placed at two sides of the metasurface to transmit and receive the electromagnetic wave.

\bibliographystyle{IEEEtran}
\bibliography{Taravati_PNAS_Ref}

\phantom{aaaaaaaaaaaaaaaaaaaaaaaaaaaaaaaaaaaaaaaaaaaaaaaaaaaaaaaaaaaaaaaaaaaaaaaaaa}
\phantom{aaaaaaaaaaaaaaaaaaaaaaaaaaaaaaaaaaaaaaaaaaaaaaaaaaaaaaaaaaaaaaaaaaaaaaaaaa}
\phantom{aaaaaaaaaaaaaaaaaaaaaaaaaaaaaaaaaaaaaaaaaaaaaaaaaaaaaaaaaaaaaaaaaaaaaaaaaa}
\phantom{aaaaaaaaaaaaaaaaaaaaaaaaaaaaaaaaaaaaaaaaaaaaaaaaaaaaaaaaaaaaaaaaaaaaaaaaaa}
\phantom{aaaaaaaaaaaaaaaaaaaaaaaaaaaaaaaaaaaaaaaaaaaaaaaaaaaaaaaaaaaaaaaaaaaaaaaaaa}
\phantom{aaaaaaaaaaaaaaaaaaaaaaaaaaaaaaaaaaaaaaaaaaaaaaaaaaaaaaaaaaaaaaaaaaaaaaaaaa}
\phantom{aaaaaaaaaaaaaaaaaaaaaaaaaaaaaaaaaaaaaaaaaaaaaaaaaaaaaaaaaaaaaaaaaaaaaaaaaa}
\phantom{aaaaaaaaaaaaaaaaaaaaaaaaaaaaaaaaaaaaaaaaaaaaaaaaaaaaaaaaaaaaaaaaaaaaaaaaaa}
\phantom{aaaaaaaaaaaaaaaaaaaaaaaaaaaaaaaaaaaaaaaaaaaaaaaaaaaaaaaaaaaaaaaaaaaaaaaaaa}
\phantom{aaaaaaaaaaaaaaaaaaaaaaaaaaaaaaaaaaaaaaaaaaaaaaaaaaaaaaaaaaaaaaaaaaaaaaaaaa}
\phantom{aaaaaaaaaaaaaaaaaaaaaaaaaaaaaaaaaaaaaaaaaaaaaaaaaaaaaaaaaaaaaaaaaaaaaaaaaa}
\phantom{aaaaaaaaaaaaaaaaaaaaaaaaaaaaaaaaaaaaaaaaaaaaaaaaaaaaaaaaaaaaaaaaaaaaaaaaaa}
\phantom{aaaaaaaaaaaaaaaaaaaaaaaaaaaaaaaaaaaaaaaaaaaaaaaaaaaaaaaaaaaaaaaaaaaaaaaaaa}
\phantom{aaaaaaaaaaaaaaaaaaaaaaaaaaaaaaaaaaaaaaaaaaaaaaaaaaaaaaaaaaaaaaaaaaaaaaaaaa}
\phantom{aaaaaaaaaaaaaaaaaaaaaaaaaaaaaaaaaaaaaaaaaaaaaaaaaaaaaaaaaaaaaaaaaaaaaaaaaa}
\phantom{aaaaaaaaaaaaaaaaaaaaaaaaaaaaaaaaaaaaaaaaaaaaaaaaaaaaaaaaaaaaaaaaaaaaaaaaaa}
\phantom{aaaaaaaaaaaaaaaaaaaaaaaaaaaaaaaaaaaaaaaaaaaaaaaaaaaaaaaaaaaaaaaaaaaaaaaaaa}
\phantom{aaaaaaaaaaaaaaaaaaaaaaaaaaaaaaaaaaaaaaaaaaaaaaaaaaaaaaaaaaaaaaaaaaaaaaaaaa}
\phantom{aaaaaaaaaaaaaaaaaaaaaaaaaaaaaaaaaaaaaaaaaaaaaaaaaaaaaaaaaaaaaaaaaaaaaaaaaa}
\phantom{aaaaaaaaaaaaaaaaaaaaaaaaaaaaaaaaaaaaaaaaaaaaaaaaaaaaaaaaaaaaaaaaaaaaaaaaaa}
\phantom{aaaaaaaaaaaaaaaaaaaaaaaaaaaaaaaaaaaaaaaaaaaaaaaaaaaaaaaaaaaaaaaaaaaaaaaaaa}
\phantom{aaaaaaaaaaaaaaaaaaaaaaaaaaaaaaaaaaaaaaaaaaaaaaaaaaaaaaaaaaaaaaaaaaaaaaaaaa}
\phantom{aaaaaaaaaaaaaaaaaaaaaaaaaaaaaaaaaaaaaaaaaaaaaaaaaaaaaaaaaaaaaaaaaaaaaaaaaa}
\phantom{aaaaaaaaaaaaaaaaaaaaaaaaaaaaaaaaaaaaaaaaaaaaaaaaaaaaaaaaaaaaaaaaaaaaaaaaaa}
\phantom{aaaaaaaaaaaaaaaaaaaaaaaaaaaaaaaaaaaaaaaaaaaaaaaaaaaaaaaaaaaaaaaaaaaaaaaaaa}
\phantom{aaaaaaaaaaaaaaaaaaaaaaaaaaaaaaaaaaaaaaaaaaaaaaaaaaaaaaaaaaaaaaaaaaaaaaaaaa}
\phantom{aaaaaaaaaaaaaaaaaaaaaaaaaaaaaaaaaaaaaaaaaaaaaaaaaaaaaaaaaaaaaaaaaaaaaaaaaa}
\phantom{aaaaaaaaaaaaaaaaaaaaaaaaaaaaaaaaaaaaaaaaaaaaaaaaaaaaaaaaaaaaaaaaaaaaaaaaaa}
\phantom{aaaaaaaaaaaaaaaaaaaaaaaaaaaaaaaaaaaaaaaaaaaaaaaaaaaaaaaaaaaaaaaaaaaaaaaaaa}
\phantom{aaaaaaaaaaaaaaaaaaaaaaaaaaaaaaaaaaaaaaaaaaaaaaaaaaaaaaaaaaaaaaaaaaaaaaaaaa}
\phantom{aaaaaaaaaaaaaaaaaaaaaaaaaaaaaaaaaaaaaaaaaaaaaaaaaaaaaaaaaaaaaaaaaaaaaaaaaa}
\phantom{aaaaaaaaaaaaaaaaaaaaaaaaaaaaaaaaaaaaaaaaaaaaaaaaaaaaaaaaaaaaaaaaaaaaaaaaaa}
\phantom{aaaaaaaaaaaaaaaaaaaaaaaaaaaaaaaaaaaaaaaaaaaaaaaaaaaaaaaaaaaaaaaaaaaaaaaaaa}
\phantom{aaaaaaaaaaaaaaaaaaaaaaaaaaaaaaaaaaaaaaaaaaaaaaaaaaaaaaaaaaaaaaaaaaaaaaaaaa}
\phantom{aaaaaaaaaaaaaaaaaaaaaaaaaaaaaaaaaaaaaaaaaaaaaaaaaaaaaaaaaaaaaaaaaaaaaaaaaa}
\phantom{aaaaaaaaaaaaaaaaaaaaaaaaaaaaaaaaaaaaaaaaaaaaaaaaaaaaaaaaaaaaaaaaaaaaaaaaaa}
\phantom{aaaaaaaaaaaaaaaaaaaaaaaaaaaaaaaaaaaaaaaaaaaaaaaaaaaaaaaaaaaaaaaaaaaaaaaaaa}
\phantom{aaaaaaaaaaaaaaaaaaaaaaaaaaaaaaaaaaaaaaaaaaaaaaaaaaaaaaaaaaaaaaaaaaaaaaaaaa}
\phantom{aaaaaaaaaaaaaaaaaaaaaaaaaaaaaaaaaaaaaaaaaaaaaaaaaaaaaaaaaaaaaaaaaaaaaaaaaa}
\phantom{aaaaaaaaaaaaaaaaaaaaaaaaaaaaaaaaaaaaaaaaaaaaaaaaaaaaaaaaaaaaaaaaaaaaaaaaaa}
\phantom{aaaaaaaaaaaaaaaaaaaaaaaaaaaaaaaaaaaaaaaaaaaaaaaaaaaaaaaaaaaaaaaaaaaaaaaaaa}
\phantom{aaaaaaaaaaaaaaaaaaaaaaaaaaaaaaaaaaaaaaaaaaaaaaaaaaaaaaaaaaaaaaaaaaaaaaaaaa}
\phantom{aaaaaaaaaaaaaaaaaaaaaaaaaaaaaaaaaaaaaaaaaaaaaaaaaaaaaaaaaaaaaaaaaaaaaaaaaa}
\pagebreak
\phantom{aaaaaaaaaaaaaaaaaaaaaaaaaaaaaaaaaaaaaaaaaaaaaaaaaaaaaaaaaaaaaaaaaaaaaaaaaa}
\pagebreak

\section*{{\Large \textbf{\textsf{Supporting Information}}}}\label{sec:supp_inf}
  \renewcommand{\theequation}{S\arabic{equation}}
  \renewcommand{\thefigure}{S\arabic{figure}}
  \setcounter{equation}{0}  
  \setcounter{figure}{0}
\subsection*{\textsf{\textbf{Coupled-Structure Resonances Suppression}}}
In this section, we provide the exact analytical solution for the scattering parameters of the reciprocal and nonreciprocal unit cells in Figs.~\ref{Fig:unit_cell_T}A and~\ref{Fig:unit_cell_T}D. While the solution for the former case gives more insight into the patch and coupled-structure resonances, plotted in Fig.~\ref{Fig:unit_cell_T}C, the solution for the latter case explains how the unilateral device suppresses the coupled-structure resonances, leading to the result of Fig.~\ref{Fig:unit_cell_T}F.

\begin{figure}[!ht]
\begin{center}
\psfrag{A}[c][c][0.9]{\shortstack{region 1\\$\eta_1,\beta_1$}}
\psfrag{B}[c][c][0.9]{\shortstack{region 2\\$\eta_2,\beta_2$}}
\psfrag{C}[c][c][0.9]{\shortstack{region 3\\$\eta_3,\beta_3$}}
\psfrag{D}[c][c][0.9]{\shortstack{region 4\\$\eta_4,\beta_4$}}
\psfrag{E}[c][c][0.9]{\shortstack{region 5\\$\eta_5,\beta_5$}}
\psfrag{F}[l][c][0.8]{$T_{21}$}
\psfrag{G}[l][c][0.8]{$T_{32}$}
\psfrag{H}[l][c][0.8]{$T_{43}$}
\psfrag{I}[l][c][0.8]{$T_{54}$}
\psfrag{a}[c][c][0.8]{$d_\text{2}$}
\psfrag{b}[c][c][0.8]{$d_\text{3}$}
\psfrag{c}[c][c][0.8]{$d_\text{4}$}
\psfrag{d}[c][c][0.8]{$R_{12}$}
\psfrag{e}[l][c][0.8]{$R_{21}$}
\psfrag{f}[c][c][0.8]{$R_{23}$}
\psfrag{g}[l][c][0.8]{$R_{32}$}
\psfrag{h}[c][c][0.8]{$R_{34}$}
\psfrag{i}[l][c][0.8]{$R_{43}$}
\psfrag{j}[c][c][0.8]{$R_{45}$}
\includegraphics[width=0.9\columnwidth]{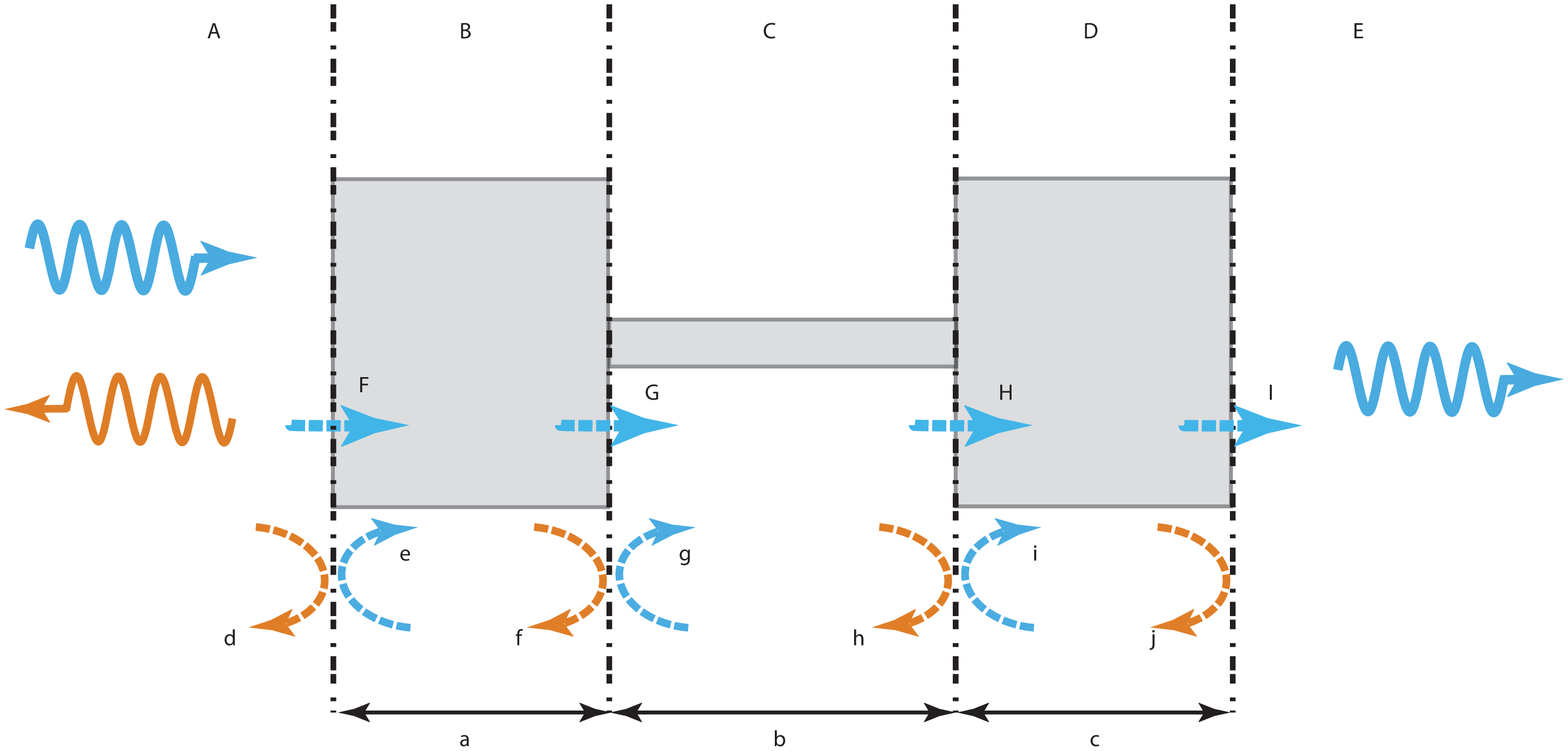}
\caption{Multiple scattering in the unfolded version of the SCS structures in Figs.~\ref{Fig:unit_cell_T}A and~\ref{Fig:unit_cell_T}D.}
\label{Fig:mult_trans}
\end{center}
\end{figure}

Figure~\ref{Fig:mult_trans} shows the general representation of the unfolded version of the SCS structures in Figs.~\ref{Fig:unit_cell_T}A and~\ref{Fig:unit_cell_T}D, where wave propagation through the SCS structure from one side to the other side of the metasurface is decomposed in five propagation regions. Since microstrip transmission lines are inhomogeneous, their wavenumbers depend on the width of the structure~\cite{Garg_2001}, and therefore the wavenumbers in different regions are different, i.e. $\beta_\text{1}(=\beta_5)\neq \beta_\text{2}(=\beta_4) \neq \beta_\text{3}$.

The total electric field in the $n$th region, where $n=1,\ldots,5$, consists of forward and backward waves as

\begin{equation}
E_n=V_n^+ e^{-j \beta_n z}+ V_n^- e^{j \beta_n z},
\label{eqa:aa}
\end{equation}

\noindent where $V_n^+$ and $V_n^-$ are the amplitudes of the forward and backward waves, respectively, and $\beta_n$ is the wavenumber. It should be noted that the backward waves, propagating along $-z$ direction, are due to reflection at the different interfaces between adjacent regions. Upon application of boundary conditions at the interface between regions $n$ and $n+1$, the total transmission and total reflection coefficients between regions $n$ and $n+1$ are found as as~\cite{Chew_1995}

\begin{subequations}
\begin{equation}
\widetilde{T}_{n+1,n} = \frac{V_{n+1}^+}{V_{n}^+}=\frac{T_{n+1,n} e^{-j (\beta_{n}-\beta_{n+1}) z}}{1-R_{n+1,n} \widetilde{R}_{n+1,n+2} e^{-j 2\beta_{n+1} d_{n+1}}},
\label{eqa:xxx}
\end{equation}
\begin{equation}
\widetilde{R}_{n,n+1} = \frac{R_{n,n+1}+\widetilde{R}_{n+1,n+2} e^{-j 2\beta_{n+1} d_{n+1}}}{1+R_{n,n+1} \widetilde{R}_{n+1,n+2} e^{-j 2\beta_{n+1} d_{n+1}}}.
\label{eqa:www}
\end{equation}
\label{eqa:TRtotal}
\end{subequations}

\noindent where $R_{n,n+1}=(\eta_{n+1}-\eta_{n})/(\eta_{n+1}+\eta_{n})$, with $\eta_n$ being the intrinsic impedance of region~$n$, is the local reflection coefficient within region $n$ between regions $n$ and $n+1$, and $R_{n+1,n}=-R_{n,n+1}$. The local transmission coefficient from region $n$ to region $n+1$ is then found as $T_{n+1,n}=1+R_{n,n+1}$.

The factor $e^{-j (\beta_{n}-\beta_{n+1}) z}$ in~\eqref{eqa:xxx} shows that, due to the nonuniformity of structure in Fig.~\ref{Fig:mult_trans}, a phase shift corresponding to the difference between the wavenumbers in adjacent regions occurs at each interface. The total transmission from region~1 to region~$N$ is the product of the transmissions from all interfaces and phase shift inside each region

\begin{equation}
s_\text{N,1} = \prod_{n=1}^{N-1} \widetilde{T}_{n+1,n} e^{-j \beta_n d_n}.
\label{eqa:mm}
\end{equation}

\begin{figure}[!ht]
\begin{center}
\includegraphics[width=0.9\columnwidth]{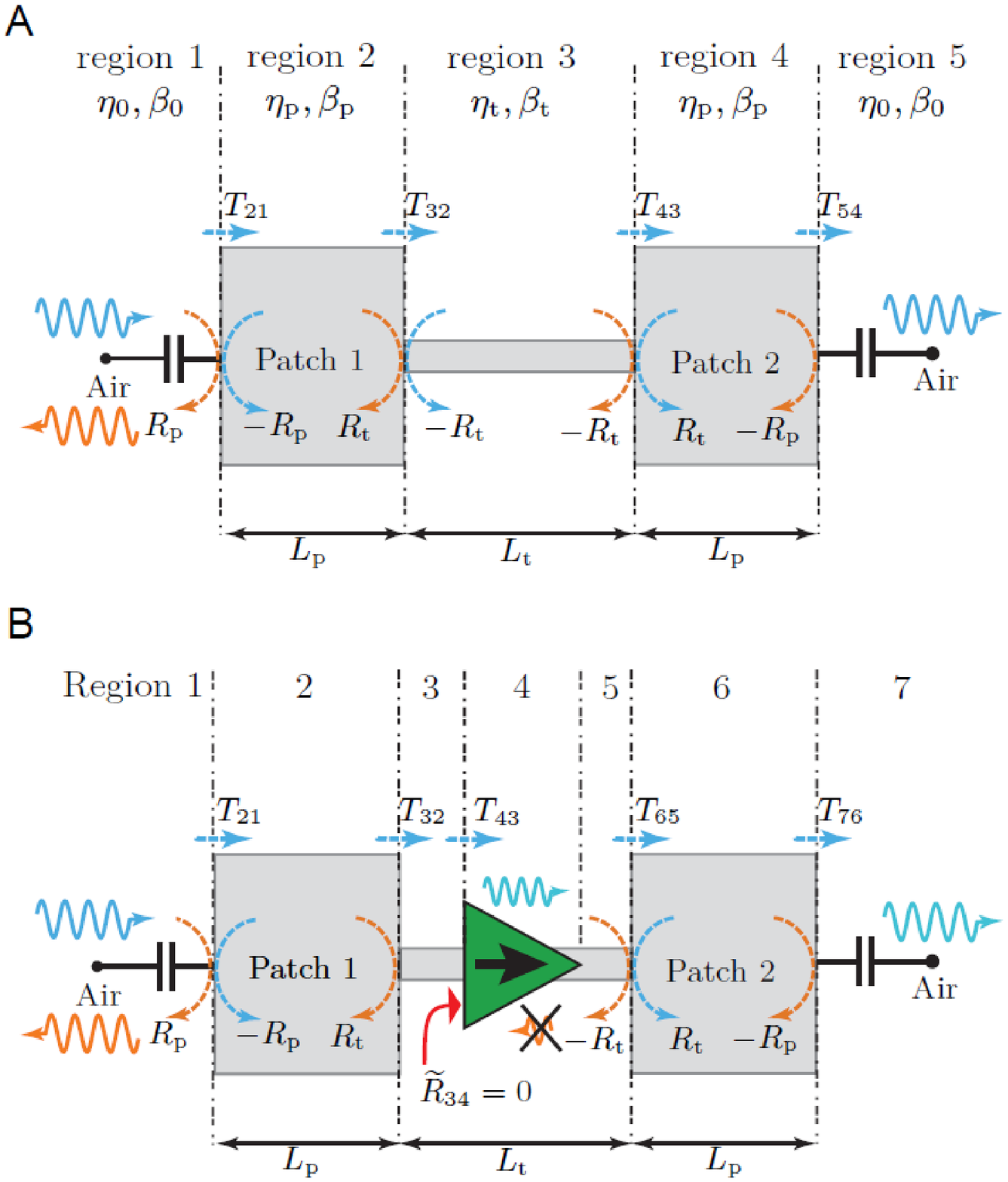}
\caption{Wave interference explanation of the responses in Fig.~\ref{Fig:unit_cell_T}. (A)~Reciprocal case (Figs.~\ref{Fig:unit_cell_T}A,~\ref{Fig:unit_cell_T}B,~\ref{Fig:unit_cell_T}C). (B)~Nonreciprocal case (Figs.~\ref{Fig:unit_cell_T}D,~\ref{Fig:unit_cell_T}E,~\ref{Fig:unit_cell_T}F).}
\label{Fig:eq_circuit}
\end{center}
\end{figure}

Figure~\ref{Fig:eq_circuit}A shows the unfolded version of the SCS architecture in Fig.~\ref{Fig:unit_cell_T}A where, comparing with the general representation of the problem in Fig.~\ref{Fig:mult_trans}, we denote ${\beta_1=\beta_5=\beta_0}$, ${\beta_2=\beta_4=\beta_\text{p}}$ and    ${\beta_3=\beta_\text{t}}$ the wavenumbers in the air, in the two patches, and in the interconnecting transmission line, respectively. We subsequently denote ${R_{1,2}=-R_{2,1}=-R_{4,5}=R_\text{p}=(\eta_\text{p}-\eta_0)/ (\eta_\text{p}+\eta_0)}$ the reflection coefficient at the interface between a patch and the air, and ${R_{2,3}=-R_{3,2}=-R_{3,4}=R_{4,3}=R_\text{t}=(\eta_\text{t}-\eta_\text{p})/(\eta_\text{t}+\eta_\text{p})}$ the local reflection coefficient at the interface between a patch and the interconnecting transmission line.

The total transmission coefficient for the reciprocal SCS metasurface of Fig.~\ref{Fig:eq_circuit}A, from region 1 to region 5, reads then

\begin{equation}
S_\text{21,Rec}=s_\text{5,1} = \prod_{n=1}^{4} \widetilde{T}_{n+1,n} e^{-j \beta_n d_n},
\label{eqa:mm2}
\end{equation}

\noindent where $\widetilde{T}_{n+1,n}$, for $n=1,\ldots,4$ is provided by~\eqref{eqa:xxx} with ~\eqref{eqa:www}. In particular,

\begin{equation}
\begin{split}
\widetilde{R}_\text{2,3,Rec} &= \frac{R_\text{t}+\widetilde{R}_{3,4} e^{-j 2\beta_\text{t} d_\text{t}}}{1+R_\text{t} \widetilde{R}_{3,4} e^{-j 2\beta_\text{t} d_\text{t}}}\\
&= \frac{R_\text{t}+R_\text{t}^2 R_\text{p} e^{-j 2\beta_{p} d_{p}}- (R_\text{t}+ R_\text{p} e^{-j 2\beta_{p} d_{p}})e^{-j 2\beta_{t} d_{t}}}{1+  R_\text{t} R_\text{p} e^{-j 2\beta_{p} d_{p}} -  R_\text{t} (R_\text{t}+ R_\text{p} e^{-j 2\beta_{p} d_{p}}) e^{-j 2\beta_{t} d_{t}}}
\end{split}
\label{eqa:jjj}
\end{equation}

\noindent will be used later.

After some algebraic manipulations in~\eqref{eqa:mm2}, the total transmission coefficient from the reciprocal SCS structure in Fig.~\ref{Fig:eq_circuit}A is found in terms of local reflection coefficients as

\begin{equation}
S_\text{21,Rec}= \frac{(1-R_\text{p}^2) (1-R_\text{t}^2) e^{j (\beta_\text{p}+ \beta_\text{0}-2\beta_\text{t}) d_\text{t} }  }{ (R_\text{p}R_\text{t}+e^{j 2\beta_\text{p} d_\text{p} } )^2 -  (R_\text{t} e^{j 2\beta_\text{p} d_\text{p}}+R_\text{p} )^2 e^{-j 2\beta_\text{t} d_\text{t}}}.
\label{eqa:nn_Rec}
\end{equation}

\noindent The term $e^{-j 2\beta_\text{t} d_\text{t}}$ in the denominator of this expression corresponds to the round-trip propagation through the middle transmission line, whose multiplication by $e^{j 4\beta_\text{p} d_\text{p} }$ in the adjacent bracket corresponds to the patch-line-patch coupled-structure resonance, with length $2d_\text{p}+d_\text{t}$.

Figure~\ref{Fig:eq_circuit}B shows the unfolded version of the nonreciprocal SCS structure in Fig.~\ref{Fig:unit_cell_T}D, where a unilateral device is placed at the middle of the interconnecting transmission line. Note that in this case the structure is decomposed in 7 (as opposed to 5) regions, with extra parameters straightforwardly following from the reciprocal case.

Assuming that the input and output ports of the unilateral device are matched, i.e. $R_{34}=R_{54}=0$, then one has $\widetilde{R}_{34}=0$ and the backward wave is completely absorbed by the device, i.e. $T_{34}=\widetilde{T}_{34}=0$ while the forward wave is amplified by the device as $T_{43}=\widetilde{T}_{43}=G$. Then, the total reflection coefficients at the interface between regions 2 and 3, given by~\eqref{eqa:jjj} in the reciprocal case, reduces to

\begin{equation}
\widetilde{R}_\text{2,3,NR} = R_\text{t}.
\label{eqa:ppp}
\end{equation}

\noindent This relation, compared with the one for the reciprocal case, reveals the suppression of the multiple reflections in the interconnecting transmission line. The total transmission coefficient from the nonreciprocal SCS structure may be found as

\begin{equation}
S_\text{21,NRec}= \frac{ G (1-R_\text{p}^2) (1-R_\text{t}^2) e^{-j (\beta_\text{t} d_\text{t} -2 \beta_\text{0}d_\text{p} ) }  }{ (R_\text{p}R_\text{t} +e^{j 2\beta_\text{p} d_\text{p} })^2 }.
\label{eqa:nn_NR}
\end{equation}

Comparing the denominator of~\eqref{eqa:nn_NR} with that of the reciprocal case in~\eqref{eqa:nn_Rec}, shows that the coupled-structure resonances, corresponding to the second term of the denominator, have disappeared due to the suppression of the multiple reflections in the middle transmission line, restricting the spectrum to the harmonic resonances of the two patches, amplified by $G$.

\begin{figure}[!ht]
\begin{center}
\includegraphics[width=1\columnwidth]{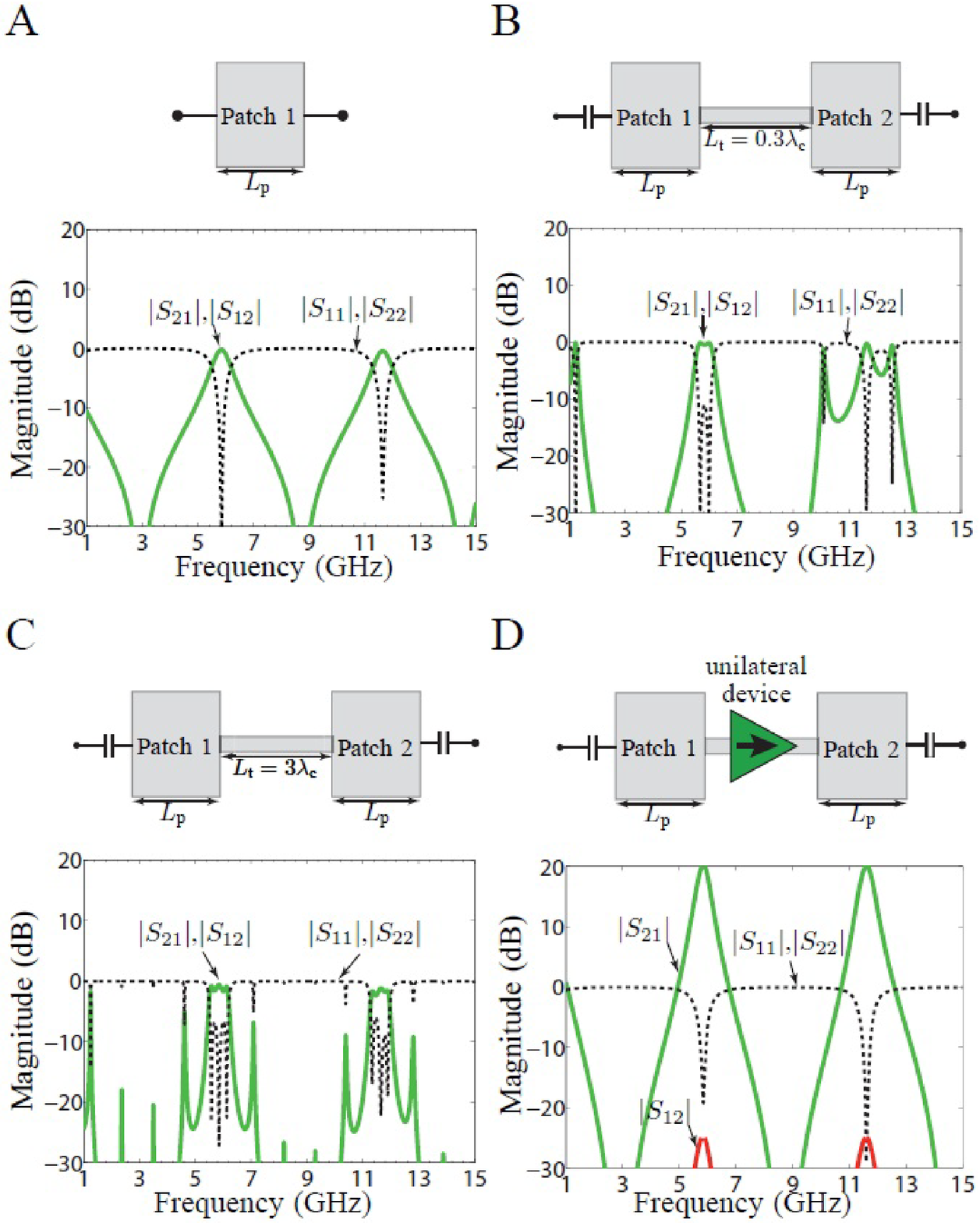}
\caption{Scattering parameter frequency responses of the structures in Fig.~\ref{Fig:eq_circuit}. (A)~Isolated patch. (B)~Structure (reciprocal) in Fig.~\ref{Fig:eq_circuit}A with $L_\text{t}=0.3\lambda_\text{c}$. (C)~Same structure (reciprocal) as in (B) except for $L_\text{t}=3\lambda_\text{c}$ (D)~Structure in Fig.~\ref{Fig:eq_circuit}B (nonreciprocal) still with $L_\text{t}=3\lambda_\text{c}$.}
\label{Fig:eq_s2p}
\end{center}
\end{figure}

Figure~\ref{Fig:eq_s2p}A shows the magnitude of the scattering parameters for a single isolated patch, where transmission ($|S_{21}|=1$) occurs at the harmonic resonance frequencies of the patch, $nf_0$ ($n$ integer), where $L_\text{p}=n \lambda_\text{p}/2=n\lambda_0/(2\sqrt{\epsilon_\text{eff}})$ with $\epsilon_\text{eff}$ being the effective permittivity~\cite{Garg_2001}. Figure~\ref{Fig:eq_s2p}B plots the magnitude of the scattering parameters of the coupled structure formed by the two patches interconnected by a short transmission line of $L_\text{t}=0.3\lambda_\text{0}$, given by~\eqref{eqa:nn_Rec}. We see that, in addition to the single patch resonances at $f=nv_\text{p}/(2L_\text{p})$, extra resonances appear in the spectrum, corresponding to the aforementioned coupled-structure resonances. Increasing the length of the interconnecting transmission line to $L_\text{t}=3\lambda_\text{0}$ yields the results presented in Fig.~\ref{Fig:eq_s2p}C. As expected, more coupled-structure resonances appear in the response due to the longer electrical length of the overall structure, while the patch resonances remain fixed. Finally, we place the unilateral device in the middle of the interconnecting transmission line, still with $L_\text{t}=3\lambda_\text{0}$. Figure~\ref{Fig:eq_s2p}D shows the corresponding scattering parameters, where all the coupled resonances in Fig.~\ref{Fig:eq_s2p}C have been completely suppressed due to the absorbtion of multiple reflections from the patches by the unilateral device. It should be noted that the forward amplification, $|S_{21}|>1$, and backward isolation, $|S_{12}|\ll 1$, are due to the nonreciprocal amplification of the unilateral device.

\end{document}